\documentclass[11pt]{aastex}
\usepackage{graphicx,emulateapj5,apjfonts}
\usepackage{onecolfloat}
\usepackage{epsf}

\shortauthors{Bell et al.}
\shorttitle{Star formation and the growth of stellar mass}

\newcommand{\peg}{{\sc P\'egase }}
\newcommand{\combo}{{COMBO-17} }

\slugcomment{{\sc The Astrophysical Journal, in press} }

\begin{document}


\def\head{

\title{Star formation and the growth of stellar mass }

\author{Eric F.\ Bell$^1$, Xian Zhong Zheng$^{1,2}$, Casey Papovich$^3$, Andrea Borch$^4$, 
Christian Wolf$^5$ and 
Klaus Meisenheimer$^1$}
\affil{$^1$ Max-Planck-Institut f\"ur Astronomie,
K\"onigstuhl 17, D-69117 Heidelberg, Germany; \texttt{bell@mpia.de}\\
$^2$  Purple Mountain Observatory, CAS, West Beijing Road 2, 
Nanjing, 210008, P.\ R.\ China \\
$^3$  Steward Observatory, The University of Arizona, 933 North
Cherry Avenue, Tucson, AZ 85721, USA \\
$^4$ Astronomisches Rechen-Institut, M\"onchhofstr. 12-14, D69120 Heidelberg, 
Germany \\
$^5$ Department of Physics, Denys Wilkinson Bldg., University
of Oxford, Keble Road, Oxford, OX1 3RH, UK \\
}

\begin{abstract}

Recent observations have demonstrated a significant growth in 
the integrated stellar mass of the red sequence since $z=1$, 
dominated by a steadily increasing number of galaxies with 
stellar masses $M_* < 10^{11} M_{\sun}$.  
In this paper, we use the COMBO-17 photometric redshift survey
in conjunction with deep Spitzer 24{\micron} data to explore
the relationship between star formation and the growth of stellar
mass.
We calculate `star formation rate functions' in four
different redshift slices, splitting also into contributions
from the red sequence and blue cloud for the first time.  
We find that the growth of stellar mass since $z=1$ is
consistent with the integrated star formation rate.  Yet, 
most of the stars formed are in blue cloud galaxies. If 
the stellar mass already in, and formed in, $z<1$ blue cloud 
galaxies were to stay in the blue cloud the total stellar
mass in blue galaxies would be dramatically overproduced.
We explore the expected evolution of stellar mass functions, 
finding that in this picture the number of massive 
$M_* > 3 \times 10^{10} M_{\sun}$ blue galaxies would also be 
overproduced; i.e., most of the 
new stars formed in blue cloud galaxies are in the 
massive galaxies.  We explore a simple truncation 
scenario in which these `extra' blue galaxies have 
their star formation suppressed by an unspecified mechanism or mechanisms;
simple cessation of star formation in these extra blue
galaxies is approximately sufficient to build up the red sequence
at $M_* \la 10^{11}M_{\sun}$.  

\end{abstract}

\keywords{galaxies: general ---  
galaxies: evolution --- galaxies: stellar content ---
infrared: galaxies }
}

\twocolumn[\head]

\section{Introduction}
Understanding where and when the stars in galaxies form is
one of the key goals of observational extragalactic astronomy.
In the last decade, magnificent progress has been 
made towards this goal, with the construction of
large galaxy redshift surveys allowing the evolution 
of cosmic star formation rate \citep[e.g.,][]{madau96,lilly96,steidel99,flores99,haarsma00,hopkins04,lefloch05} and the build-up of 
stellar mass \citep[e.g.,][]{brinchmann00,cole01,dickinson03,rudnick03,rudnick06,fontana,fontana06,drory04,drory05,borch} to be estimated.  Overall, if one assumes an 
stellar initial mass function with a power-law slope
similar to a \citet{salp} value for masses $> 1 M_{\sun}$,
which is universally-applicable on galaxy-wide scales 
\citep[see, e.g.,][for a critical discussion of this issue]{elmegreen06},
the evolution of total stellar mass in the universe
is reasonably well-reproduced by the integrated cosmic-averaged star
formation rate (SFR), provided that one accounts for stellar
mass loss \citep{cole01,rudnick03,borch}.  

It has also become clear that there is a bimodal distribution 
of galaxy colors at all $z < 1$, with a relatively
narrow red sequence dominated by non-star-forming galaxies
and a blue cloud of star-forming galaxies 
\citep[see Fig.\ \ref{fig:data}]{strateva01,blanton03,bell04,willmer06}.  
More recently, is has been demonstrated that the stellar mass
on the red sequence has built up by a substantial amount --- 
at least 50\% of the stellar mass 
in present-day red sequence galaxies has come into place
since $z \sim 1$ \citep{chen03,bell04,faber06,borch,brown06}. 
Given that relatively few stars form in red-sequence
galaxies, this build-up has been argued to be driven by the global
suppression of star formation in some fraction
of previously blue, star-forming galaxies through 
a variety of possible physical processes, e.g., 
galaxy merging, the suppression of star formation in 
dense environments, or gas consumption by star-forming disks. 

Yet, this qualitative picture leaves many questions unanswered.
What is the contribution of dust-obscured star formation 
in red galaxies to the growth of stellar mass on the 
red sequence?  Can simple truncation of star formation 
in some fraction of blue galaxies provide enough stellar mass to feed
the growth of the red sequence?  The `average' disk galaxy is several
times less massive than typical red sequence galaxies; is it possible to create
early-type galaxies through truncation of star formation in the
presumably rather low-mass blue cloud population?

In this paper, we present a first, crude attempt at addressing
some of these questions.  We combine color-split stellar
mass functions \citep{borch} from the COMBO-17 photometric redshift
survey \citep{wolf03,wolf04} with SFR functions derived
from ultraviolet (UV) and infrared (IR) luminosities of galaxies in 
COMBO-17 \citep[see, e.g.,][for IR luminosity functions derived using spectroscopic redshifts and COMBO-17 redshifts]{lefloch05}.  While both 
stellar masses and SFRs are model-dependent, depend on the 
assumption of a universally-applicable galaxy-scale stellar
initial mass function (IMF), and suffer from 
considerable systematic and random uncertainties, such an 
analysis allows one to explore the growth of 
stellar mass with cosmic time and attempt to understand
in which types of galaxies the bulk of stellar mass forms, 
and in which types of galaxies most stellar mass ends
up.  While necessarily qualitatitive, such an approach gives
interesting insight into some of the basic features of 
the physical processes which drive galaxy evolution.

The plan of this paper is as follows.  In \S \ref{data},
we describe the \combo redshift survey data, the construction 
of stellar masses, and the estimation of SFR functions using
Spitzer and COMBO-17 data; this section includes
the first estimate of a key observational
constraint on this analysis --- the estimation of  
SFR functions split by galaxy color, as a function of cosmic epoch.  
In \S \ref{integ}, we explore the 
growth of the total stellar mass densities in red and blue galaxies, 
and how this growth relates to observed star formation in 
the red and blue galaxy populations.  In \S \ref{diff}, 
we explore the evolution of the stellar mass functions of red 
and blue galaxies, and explore how one might expect the
stellar mass function to evolve under various simplistic 
assumptions.  These results are discussed 
in \S \ref{disc}.  In this paper, we adopt a 
$H_{0}=70$\,km\,s$^{-1}$\,Mpc$^{-1}$,
$\Omega_\Lambda=0.7$ and $\Omega_{\rm m}=0.3$ cosmology,
and assume that the distribution of stellar masses formed
follows a \citet{chabrier03} stellar IMF, averaged over galaxy
scales; such a stellar IMF gives stellar masses and SFRs 
consistent with a \citet{kroupa01}
IMF to within 10\%.

\section{The Data} \label{data}

\subsection{COMBO-17 and stellar masses} \label{sec:mass}

To date, \combo has fully surveyed three disjoint
$\sim 34' \times 33'$ southern and equatorial fields
to deep limits in 5 broad and 12 medium passbands.
Using these deep data in conjunction with 
non-evolving galaxy, star, and AGN template spectra, objects
are classified and redshifts assigned for $\sim 99$\% of the
objects to a limit of $m_R \sim 23.5$.  Typical galaxy redshift accuracy
is $\delta z/(1+z) \sim 0.02$ \citep{wolf04},
allowing construction of $\sim 0.1$ mag accurate 
rest-frame colors and absolute magnitudes (accounting for distance
and $k$-correction uncertainties).  Astrometric accuracy is 
$\sim 0.1$\arcsec.

\citet{borch} estimated the stellar mass of galaxies in 
\combo using the 17-passband photometry in conjunction with 
a non-evolving template library derived using the 
\peg stellar population model \citep[see][for a description
of an earlier version of the model]{fioc97}.  The masses were
derived using a \citet{kroupa93} stellar IMF; the use of a
\citet{kroupa01} or \citet{chabrier03} IMF would have yielded 
stellar masses to within $\sim 10$\%.  Such masses
are quantitatively consistent with those derived using a simple
color-stellar M/L relation \citep{bell03}, and comparison of 
stellar and dynamical masses for a few $z \sim 1$ early-type
galaxies yielded consistent results to within their 
combined errors \citep[see][for more details]{borch}.
Random stellar mass errors are $< 0.3$\,dex on a galaxy-by-galaxy
basis, and systematic errors in the stellar
masses (setting the overall mass scale and its redshift evolution)
were argued to be at the 0.1\,dex level for galaxies without
ongoing or recent major starbursts; for galaxies with strong bursts, 
masses could be overestimated by $\la 0.5$\,dex.  In particular, 
the masses are derived using template stellar populations 
with an age/metallicity combination equivalent to roughly solar metallicity
and $\sim 6$\,Gyr since the start of star formation \citep{borch}; 
the templates do not evolve with redshift.  While such a 
template set is appropriate for galaxies at $z \sim 0.5-0.7$ (assuming that 
star formation starts at $z > 3$ for most massive galaxies), galaxies at
lower (higher) redshifts will have older (younger) underlying populations.
This leads to small, second-order offsets in stellar M/L at a given
`color'\footnote{The first order reddening of the stellar populations
as the Universe ages is accounted for as the aged galaxy will be fit
by a redder template.  This second order effect comes about because
the older, lower redshift galaxies will have a somewhat larger underlying
old population (carrying more mass for a given amount of light) 
than its equivalent
galaxy with similar colors at high redshift/earlier times.}.
As discussed in \citet{belldejong} and \citet[see their Fig.\ 6]{borch}, 
we expect that in our case this would lead to a slight 
underestimate (overestimate) of stellar masses at
lower (higher) redshifts at the $\sim 10-20\%$ level.  This
has the effect that the internal evolution of mass density within COMBO-17
will be artificially flattened.  We have chosen not to apply a correction 
for this oversimplification in what follows to remain consistent with 
\citet{borch}.

\begin{figure*}[th]
\begin{center}
\epsfxsize 18.0cm
\epsfbox{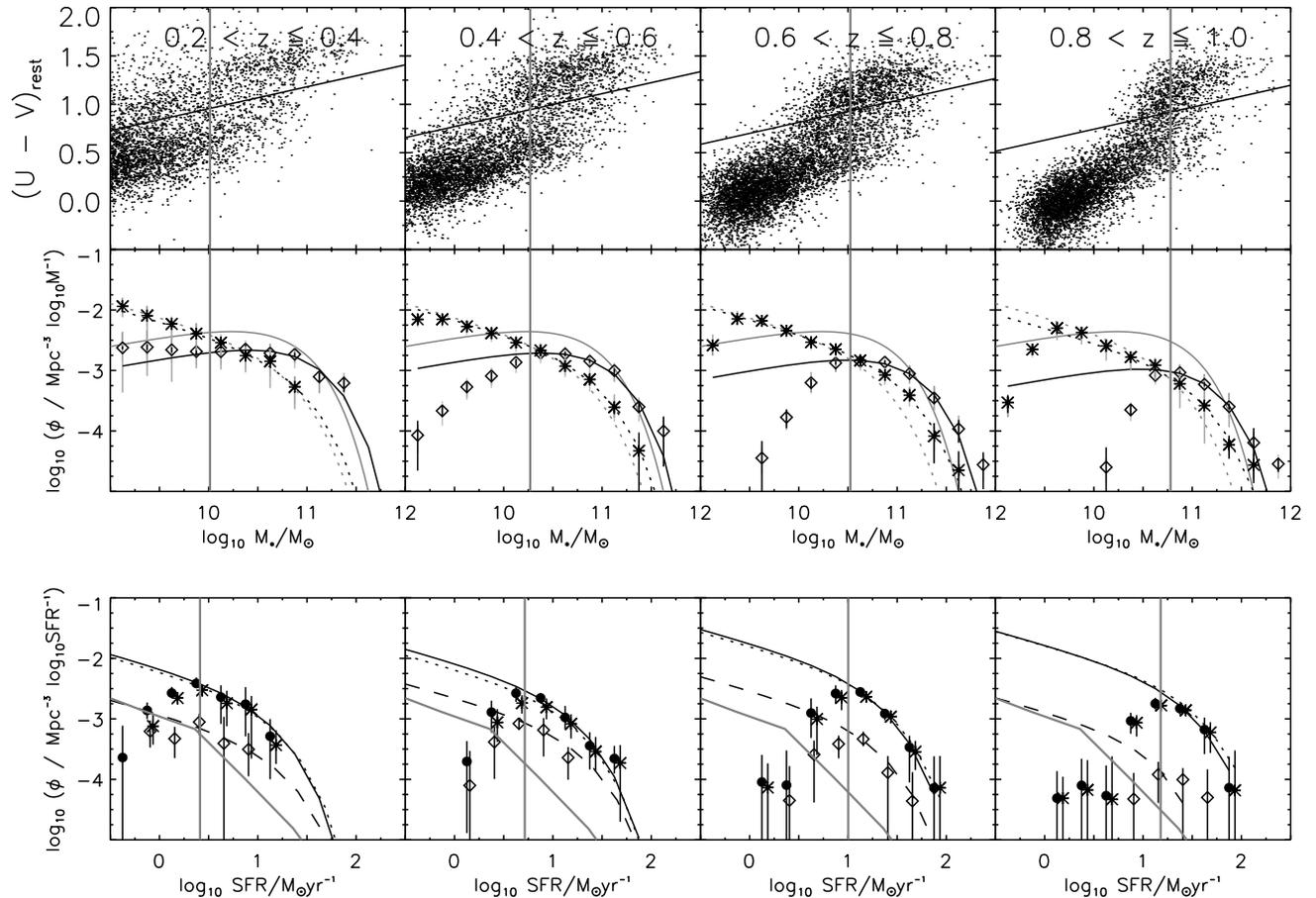}
\end{center}
\caption{\label{fig:data} 
The basic ingredients of this analysis.  
{\it Top panels:} the evolution of galaxy colors and stellar
masses over the interval $0.2<z \le 1.0$.  In each panel, 
the mass limit below which the stellar masses become incomplete is 
denoted by a thick grey line (the limit is calculated at the 
highest redshift in each redshift shell).  The solid black 
line shows the cut used to separate 
red sequence and blue cloud galaxies.
{\it Central panels:} The evolution of the stellar mass 
function.  The stellar mass function in the red sequence
(diamonds and solid line) and blue cloud (asterisks and dotted line)
is shown at each redshift interval of interest.  The $z=0$
color-split stellar mass functions are shown in each panel in gray.
Uncertainties from counting statistics are shown in black; uncertainties
from field-to-field variations are shown in gray.
{\it Lower panels:} The evolution of the SFR function.
Filled circles and solid lines denote the SFR function for
all galaxies, asterisks and dotted lines blue cloud galaxies, and 
diamonds/dashed lines red-sequence galaxies (the blue cloud
and red sequence galaxies are offset horizontally by a small 
amount for clarity).  The vertical
grey line shows the SFR limit, and the $z=0$ IR-derived SFR 
function is shown in gray in each panel for reference.
Uncertainties are determined from the difference between 
our two fields with 24{\micron} data, and our best estimate is the average
of the two fields.
}
\end{figure*}

Stellar mass functions were derived for COMBO-17 in 
four redshift shells: $0.2<z\le 0.4$, $0.4<z\le0.6$,
$0.6<z \le 0.8$, and $0.8<z \le 1$.  Mass-limited subsamples
were constructed in each redshift interval: the limiting 
mass at each redshift was: $\sim 1,1.8,3,6 \times 10^{10} {\rm M}_{\sun}$ 
\citep[see][for details of the calculation and 
discussion of the limiting stellar masses]{borch}.
Mass functions (i.e., the comoving space density of
galaxies as a function of their stellar mass) were calculated
for galaxies above this limiting mass (i.e., no $V_{max}$
corrections were applied, as the sample is 
volume-limited within this redshift shell, above
this mass limit).  Schechter functions
were fit to the mass functions above their respective
mass limits for three different cases: all galaxies,
blue cloud galaxies only, and red sequence galaxies
only.  The stellar mass functions for blue cloud galaxies 
are complete well below this formal mass limit (owing to the
low optical stellar M/L ratios of blue galaxies), therefore
the extent to which the fitted mass function follows the data points
below this mass limit is a test of the validity of our assumed faint-end
slope at each redshift of interest.
In each case, a different faint end slope was assumed:
$\alpha = -1.1, -1.45, -0.7$ for all, blue and red galaxies
respectively.  These choices were motivated by the $z \sim 0$
stellar mass functions of \citet{bell03}: we adopt their
stellar mass function for the local value in this work.
Galaxies were assigned to the red sequence using
the approximate\footnote{Owing
to slight calibration differences between the fields, the 
detailed red sequence cut is field dependent, with the 
intercept at $10^{10} M_{\sun}$ and $z=0$ 
being $U-V = $1.01, 1.06 and 1.11 in the CDFS, 
A901 and the S11 fields respectively \citep[see][for a definition of
the fields]{wolf03}.} cut $U-V \ga 1.06 -0.352z + 0.227(\log_{10} M_* - 10)$.
The results are shown in Fig.\ \ref{fig:data}.  
It is worth noting that the overall stellar mass functions 
are in excellent agreement with previous determinations
 \citep[see][]{borch}, with the advantage of tracking 
the evolution of red sequence and blue cloud galaxies 
separately.

\subsection{MIPS 24{\micron} data and star formation rates}

Spitzer has observed two of the \combo fields:
a $1{\arcdeg} \times 0{\fdg}5$ field around the extended
Chandra Deep Field South (CDFS)
in January and February 2004 as part of the time allocated to the
Spitzer Guaranteed Time Observers (GTOs), and 
a similarly-sized field around the Abell 901/902 galaxy cluster (A901)
in December 2004 and June 2005 as part of Spitzer GO-3294 (PI: Bell).
In both cases, MIPS 24{\micron} data
were taken in slow scan-map mode, with individual exposures of 10\,s.
We reduced the individual image frames using a custom data-analysis
tool (DAT) developed by the GTOs \citep{dat}.  The reduced images
were corrected for geometric distortion and combined to form full
mosaics.  The final mosaic has a pixel scale of
$1\farcs25$~pixel$^{-1}$ and an image PSF FWHM of $\simeq 6$\arcsec.
Source detection and photometry were performed using techniques
described in \citet{papovich04}; based on the analysis in that
work, we estimate that our source detection is 80\% complete at
83~$\mu$Jy in the 24{\micron} image in both the CDFS and A901
for a total exposure of $\sim 1400$\,s\,pix$^{-1}$.

In order to interpret the observed 24{\micron} emission, we
must match the 24{\micron} sources to galaxies for which 
we have redshift estimates from COMBO-17.   In both the CDFS
and A901 fields, we adopt a 1$''$ matching radius.
In the areas of the CDFS (A901) fields where there is overlap between 
the COMBO-17 redshift data and the full-depth MIPS mosaic, 
there are a total of 3255 (4251) 24{\micron} sources
with fluxes in excess of 83$\mu$Jy. 
In both fields, 70\% of the 24{\micron} 
sources with fluxes $> 83 \mu$Jy are detected by COMBO-17 in 
at least the deep $R$-band, with $m_R \la 26$.   
Some 35\% of the 24{\micron} sources have bright $m_{R,ap} < 24$
and have photometric redshift $z<1$; these 35\% of sources
contain nearly half of the total 24{\micron} flux in objects
brighter than 83$\mu$Jy.  Sources fainter than 
$m_R \ga 24$ contain the other half of the $f_{24} > 83\mu$Jy
24{\micron} sources; investigation of COMBO-17 lower confidence
photometric redshifts, their optical colors, and results from 
other studies lends weight to the argument that essentially 
all of these sources are at $z>0.8$, with the bulk lying at
$z > 1$ (e.g., Le Floc'h et al.\ 2004; Papovich et al.\ 2004; 
see Le Floc'h et al.\ 2005 for a further discussion
of the completeness of redshift information in the CDFS COMBO-17 data).  
Given the low-confidence COMBO-17 redshifts
in hand, we estimate at $\la 0.2$\,dex incompleteness in this $0.8<z<1$ bin, 
and negligible incompleteness in the $z<0.8$ bins.

The goal of our analysis of the 24{\micron} data
is to obtain estimates of SFR which account for both
unobscured (via the UV) and dust-obscured 
star formation (via the thermal IR).  Ideally, we would
have a measure of the total thermal IR flux from 8--1000{\micron};
instead, we have an estimate of IR luminosity at
one wavelength, 24{\micron}, corresponding to rest-frame 20--12{\micron}
at the redshifts of interest $z=0.2-1$.  Yet, local IR-luminous galaxies show
a surprisingly tight correlation between 
rest-frame 12--15{\micron} luminosity and total IR luminosity
\citep[e.g.,][]{spi95,cha01,rou01,papovich02}, with a scatter
of $\sim 0.15$ dex.  Following \citet{papovich02}, we 
choose to construct total IR luminosity from the observed-frame
24{\micron} data.  We use the Sbc template 
from the \citet{dev99} SED library 
to translate observed-frame 
24{\micron} flux into the 8--1000{\micron} total IR 
luminosity\footnote{Total 8--1000{\micron} IR 
luminosities are $\sim 0.3$ dex 
higher than the 42.5--122.5{\micron} 
luminosities defined by \citet{helou88},
with an obvious dust 
temperature dependence.}; such a template is found to reproduce
well the average 70{\micron} and 160{\micron} fluxes of 
$z \sim 0.7$ galaxies with $L_{IR} \ga 10^{11}L_{\sun}$ 
\citep{zheng07_70160}.
The IR luminosity uncertainties are
primarily systematic.  Firstly, there is a natural 
diversity of IR spectral shapes at a given galaxy 
IR luminosity, stellar mass, etc.; one can crudely estimate
the scale of this uncertainty by using the full
range of templates from \citet{dev99}, or by using 
templates from, e.g., \citet{dale01} instead.  This 
uncertainty is $\la$0.3\,dex.   Secondly, it is possible
that a significant fraction of $0.2<z<1.0$ galaxies have 
IR spectral energy distributions not represented
in the local Universe: while it is impossible to 
quantify this error until the advent of Herschel Space Telescope, current
results suggest that the bulk of intermediate--high redshift
galaxies have IR spectra similar to galaxies in the local
universe \citep{appleton04,elbaz05,yan05,zheng07_70160}.

We estimate SFRs
using the combined directly-observed UV light from young 
stars and the dust-reprocessed IR emission of the sample galaxies
\citep[e.g.,][]{fluxrat}.  Following \cite{bell05}, 
we estimate SFR $\psi$ using a calibration 
derived from \peg assuming a 100\,Myr-old stellar population 
with constant SFR and a \citet{chabrier03} IMF: 
\begin{equation}
\psi / ({\rm M_{\sun}\,yr^{-1}}) = 9.8 \times 10^{-11} \times
       (L_{\rm IR} + 2.2L_{\rm UV}),  \label{eqn:sfr}
\end{equation}
where $L_{\rm IR}$ is the total IR luminosity (as estimated above)
and $L_{\rm UV} = 1.5 \nu l_{\nu,2800}$ is a rough estimate
of the total integrated 1216{\AA}--3000{\AA} UV luminosity, derived using
the 2800{\AA} rest-frame luminosity from COMBO-17 $l_{\nu,2800}$.
The factor of 1.5 in the 2800{\AA}-to-total UV conversion accounts
for the UV spectral shape of a 100 Myr-old population with constant
SFR, and the UV flux is multiplied by a factor of 2.2 before being 
added to the IR luminosity to account
for light emitted longwards of 3000{\AA} and shortwards of 1216{\AA} by 
the unobscured young stars.  This SFR calibration is 
derived using identical assumptions to \citet{k98},
and the calibration is consistent with his to within 30\%
once different IMFs are accounted for.  Uncertainties in these
SFR estimates are no less than a factor of two in a galaxy-by-galaxy
sense, and systematic uncertainty in the overall SFR scale is 
likely to be less than a factor of two 
\citep[see, e.g.,][for further discussion of uncertainties]{bellsfr,bell05}.

Under the assumption that the bulk of 24{\micron} emission 
comes from star formation, not AGN activity\footnote{Less than
15\% of the total 24{\micron} emission at $z<1$ is in 
X-ray luminous AGN \citep[e.g.,][]{silva04,bell05,fran05,brand06}.  
Although some Compton-thick AGN will
be missed in such a census \citep[e.g.,][]{donley05,martinez06,alonso06}, 
a fraction of the 24{\micron} emission 
in galaxies with AGN is powered by star formation.  Thus, 
it is likely that $\sim 10$\% of the 24{\micron} emission 
is from AGN; such a contamination is a small uncertainty
compared to others inherent to this analysis, and does not affect 
our conclusions.} we proceed to calculate
SFR functions in a similar fashion to the stellar mass
function.   Limiting star formation rates were defined for each 
redshift bin: these are $(2.6,5.2,10.0,15.1) {\rm M_{\sun}\,yr^{-1}}$
for $z \sim (0.3,0.5,0.7,0.9)$ respectively.  
Again, the sample is volume-limited in each redshift shell down
to that limiting SFR.
The SFR functions are calculated
separately for all galaxies, blue cloud galaxies and red sequence
galaxies respectively.  In this contribution, 
we choose to adopt a Schechter function fit to the 
star formation rate function\footnote{In the local Universe, 
a Schechter function is a poor fit to the IR luminosity function (denoted
in the lower panels of Fig.\ \ref{fig:data} by a dark gray line).  
Our $0.2 \la z \la 1$ measurements have a modest dynamic range in 
SFR, and the data can be well-fit with either a Schechter function 
or a double power-law.  In this paper, we choose to adopt
a Schechter function fit with faint end slope $\alpha = -1.45$ 
in order to ensure consistent behavior and extrapolations 
of the mass functions and 
SFR functions.  Fitting with a double power-law would not affect
the results presented in this paper to within the uncertainties. }, 
adopting a faint end slope
$\alpha = -1.45$, noting that the total SFR density
is strongly dominated by the contribution of 
blue galaxies (see Fig.\ \ref{fig:data})\footnote{Schechter
functions are fit explicitly to all galaxies and the blue
galaxy populations; the red galaxy SFR function is estimated
by subtracting the contribution of the blue galaxies from all galaxies.}.
The best-fit parameters for these SFR functions
is given in Table \ref{tab:sfrf}.

\begin{table*}
\caption{Star Formation Rate Function Fits {\label{tab:sfrf}}}
\begin{center}
\begin{tabular}{lccccc}
\hline
\hline
$z_{lo}$ & $z_{hi}$ & 
$\phi^*$ & $\log_{10} \psi^*$ &
$\alpha$ & $\rho$  \\
 & &  $10^{-4}$Mpc$^{-3}\,\log_{10}M^{-1}\,yr$ & $M_{\sun}\,yr^{-1}$ & 
& $M_{\sun}\,yr^{-1}$\,Mpc$^{-3}$  \\
\hline
\multicolumn{6}{c}{All Galaxies} \\
\hline
0.2 & 0.4 & 9.8(6.0) & 1.11(27) & $-$1.45 & 0.020(8) \\
0.4 & 0.6 & 10.7(1.5) & 1.20(15) & $-$1.45 & 0.027(4) \\
0.6 & 0.8 & 21(5) & 1.29(12) & $-$1.45 & 0.065(8) \\
0.8 & 1.0 & 17(6) & 1.42(15) & $-$1.45 & 0.072(7) \\
\hline
\multicolumn{6}{c}{Blue cloud} \\
\hline
0.2 & 0.4 & 8.3(4.9) & 1.09(34) & $-$1.45 & 0.016(8) \\
0.4 & 0.6 & 8.3(2.0) & 1.16(17) & $-$1.45 & 0.020(6) \\
0.6 & 0.8 & 17(5) & 1.29(15) & $-$1.45 & 0.055(8) \\
0.8 & 1.0 & 15(6) & 1.45(15) & $-$1.45 & 0.068(7) \\
\hline
\end{tabular}
\\ 
\vspace{-1.0cm}
\tablecomments{Red galaxy SFR function `fits' were determined
in this paper as the overall SFR function minus the blue 
cloud SFR function.  This was necessary owing to the poor number
statistics of the red sequence SFR function; a direct fit
frequently gave unstable results.  Fit uncertainties include 
(rough) field-to-field uncertainties (half of the difference
between A901 and ECDFS) and formal fitting uncertainties.  
Systematic uncertainties in 24{\micron} to total IR conversion, 
and in the SFR calibration, and in the faint-end slope, are not 
included, and may amount to $\sim 0.3$\,dex in the estimated
SFR densities.  Note
that we parameterise the SFR function with a Schecter function.
A value of $H_0 = 70$\,km\,s$^{-1}$\,Mpc$^{-1}$ and a \citet{chabrier03}
IMF are assumed. 
 }
\end{center}
\end{table*}

In \S 4, we make use of average SFRs for galaxies
in given bins of stellar mass.  These average SFRs are
calculated following Zheng et al.\ (2006), and the methodology
and results are discussed in detail in \citet{zheng07_down}.  Briefly,
those galaxies in a given bin in stellar mass which 
were not individually-detected at 24{\micron} were
stacked, and a total 24{\micron} flux from individually-undetected
sources determined.
The total flux from individually-detected sources was then 
added, and the average flux calculated.  These average fluxes
were converted into an IR luminosity, combined with UV luminosities, and 
were then used to construct SFR estimates following
the above transformations.  Uncertainties in average fluxes were determined
by bootstrapping.   

\section{Comparison of integrated stellar mass and star
formation rate densities}   \label{integ}

\begin{figure}[t]
\epsfbox{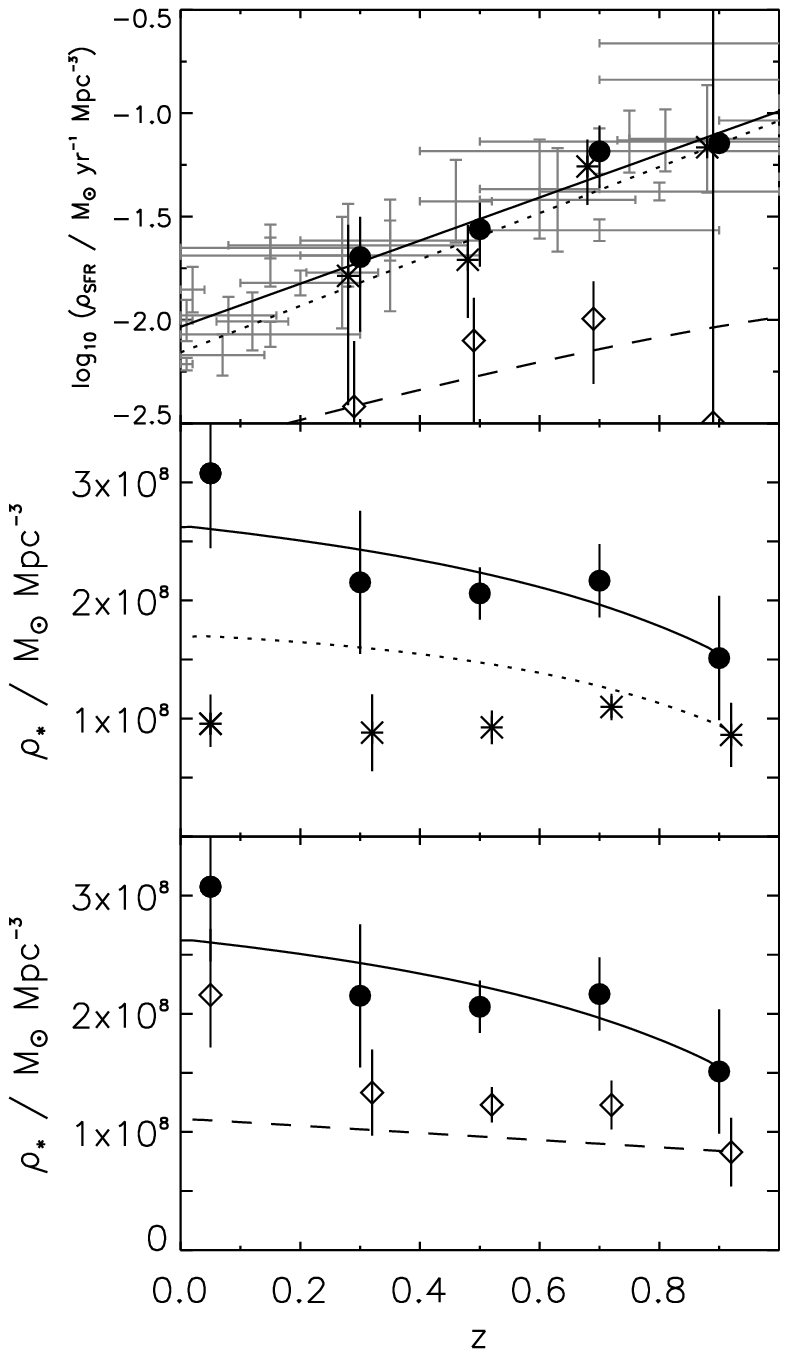}
\caption{\label{fig:integ} The cosmic evolution of 
star formation rate and stellar mass (filled circles and solid
lines), split into 
contributions from the blue cloud (asterisks and dotted lines)
and the red sequence (diamonds and dashed lines).  In the uppermost
panel, determinations of cosmic SFR are shown also in grey, 
adapted from Hopkins (2004).  In the lower panels, the data points
show the observed evolution of stellar mass as a function of redshift.
The solid line shows the predicted build-up of stellar mass, assuming
the star formation history shown by the solid line in the upper panel 
(assuming for gas recycling); it is clear that the integral of the cosmic
star formation history predicts the cosmic evolution of stellar 
mass at $z<1$ reasonably accurately.  The dotted line 
shows the predicted
evolution of total stellar mass assuming that all stars that form 
in blue galaxies remain in blue galaxies (i.e., if the blue cloud
evolves like a closed box).  The dashed line shows the corresponding 
evolution for red-sequence galaxies.  In all panels, small horizontal
offsets have been applied for clarity.
}
\end{figure}

\begin{figure}[t]
\epsfxsize=8.5cm
\epsfbox{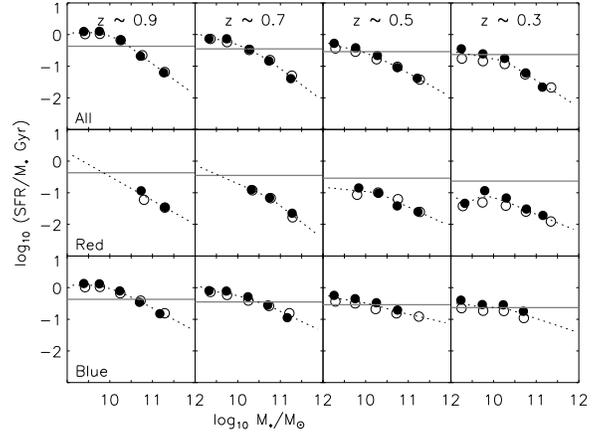}
\caption{\label{fig:spsfr} Specific SFR of $0.2<z \le 1.0$ galaxies.
Specific SFR, defined as IR-derived average SFR (from stacking) 
per unit stellar mass, is shown for all galaxies, red sequence galaxies
and blue cloud galaxies (top, middle and bottom respectively).  
Four different redshift bins are shown, and in each bin
the specific SFR is shown which corresponds to roughly constant
star formation since $z_f = 4$ (i.e., a birthrate $b$ of 1; grey 
horizontal line).  Filled points show specific SFRs as determined
for the CDFS; open points show values determined from A901.
The average blue cloud galaxy has
$b \sim 1$, red galaxies typically values between 1/3 and 0.1.
Red sequence galaxies have significant star formation, as we have made
no attempt to weed out star-forming systems with significant dust
contents from the non-star-formers on the red sequence. 
The dotted line connects the average values of specific SFR. 
}
\end{figure}

\begin{figure}[t]
\epsfxsize=8.5cm
\epsfbox{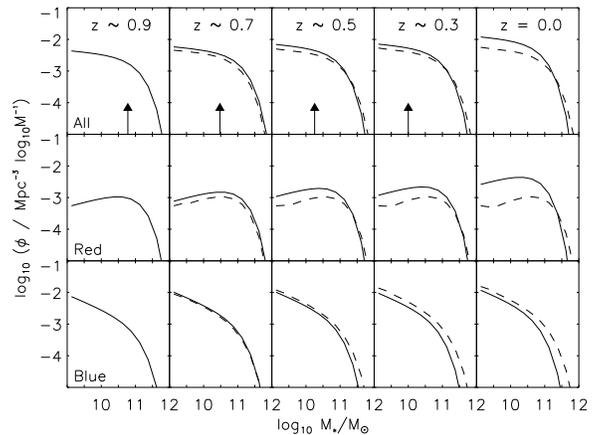}
\caption{\label{fig:hyp1} Hypothesis 1.  The evolution of the stellar mass
function for all, red and blue galaxies is shown from $z\sim 0.9$ 
to $z=0$ (solid lines in the 
top, middle and bottom rows respectively).  Dashed lines denote 
the predicted evolution of the stellar mass function, given the 
observed specific SFRs as given in Fig.\ \ref{fig:spsfr}.  In 
this case, no transfer of mass between blue and red galaxies is permitted.
Arrows denote the approximate limit above which mass functions are
well-defined.
}
\end{figure}

\begin{figure}[t]
\epsfxsize=8.5cm
\epsfbox{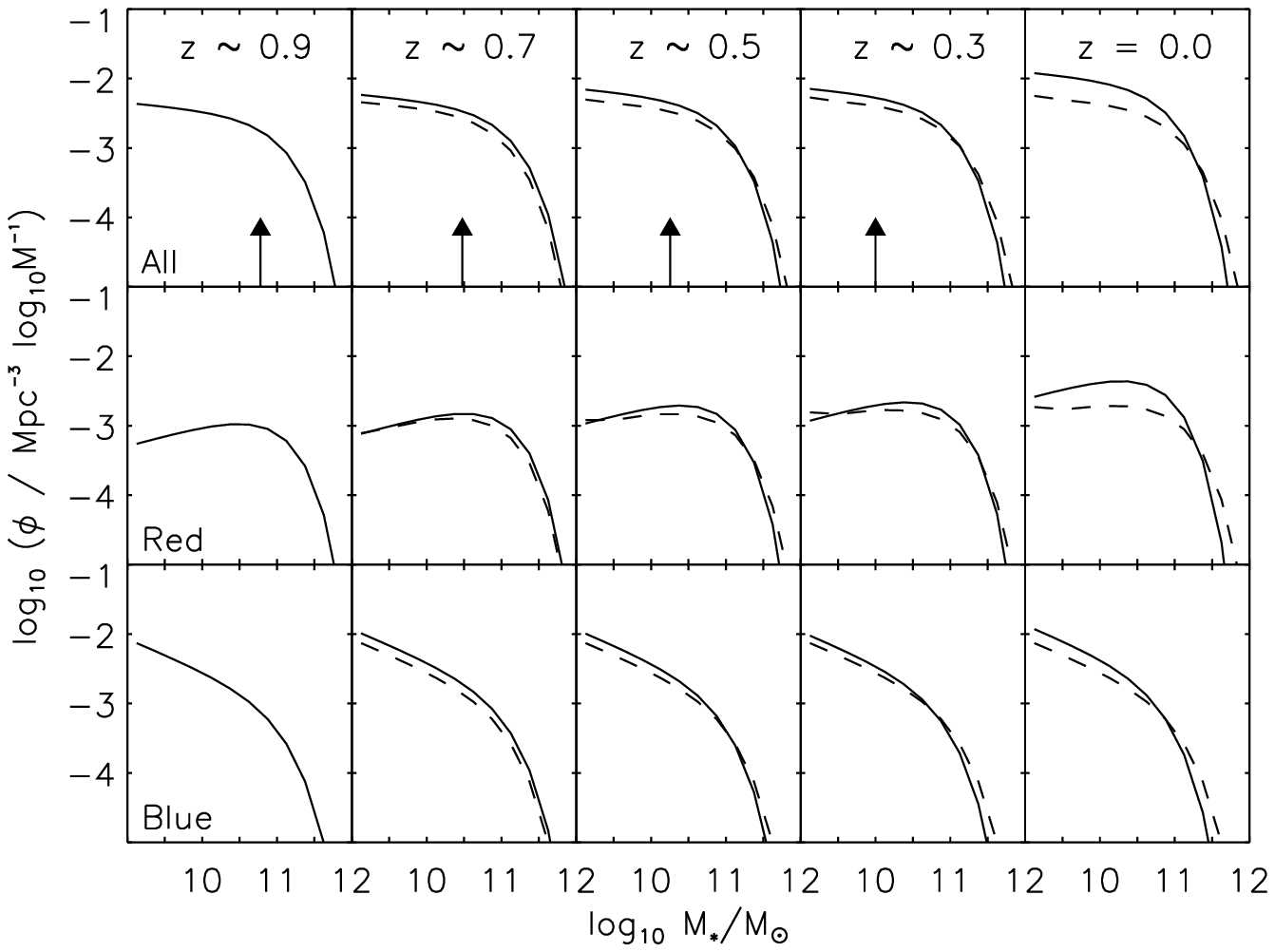}
\caption{\label{fig:hyp2} Hypothesis 2.  This figure is formatted
similarly to Fig.\ \ref{fig:hyp1}, except that in this 
case all newly-formed galaxies are assigned to the red sequence.
That is, the blue cloud is not permitted to grow, and the 
global growth in galaxy numbers at a given mass are assigned
to the red sequence.
}
\end{figure}

It is useful to establish the basic phenomenology by comparing
the evolution of the integrated stellar mass density of the 
universe with the integral of the star formation rate.
To parameterise the evolution of the total SFR, we
derive the following best fit to {\it all} total SFR 
points shown in Fig.\ \ref{fig:integ}, weighting equally (i.e., 
our measurements are included on an equal footing to literature
measurements): $\log_{10} \rho_{\rm SFR} = -2.03 + 1.04z$.  
This linear fit, shown in Fig.\ \ref{fig:integ} as the solid line, 
appears a reasonable parameterization of the data.  In order to 
track the fraction of star formation in red galaxies, we fit a linear
function to the fraction of star formation in red sequence galaxies $f_{r}$ 
as a function of redshift: $f_{r} = 0.25 -0.15z$, defined 
only over the interval $0<z\le1$.  The blue fraction $f_{b} = 1-f_{r}$.
Again, these parameterizations appear to be a reasonable description of the
data.  
  
The result of this exercise is 
shown in Fig.\ \ref{fig:integ}.  In the upper panel of 
Fig.\ \ref{fig:integ} we show the evolution of the cosmic star
formation rate; all galaxies are shown as gray error bars (adapted
from a compilation from Hopkins 2004) and solid circles (our measurements),
asterisks denote our measurements for blue cloud galaxies, and 
diamonds denote our measurements from red sequence galaxies.
In the the lower panels of 
Fig.\ \ref{fig:integ}, we show the evolution of the integrated stellar mass
density of the Universe \citep[tabulated in][solid circles]{borch}, 
along with the integral of the total star formation rate (solid line), 
accounting 
for gas recycling using the \peg stellar population synthesis
code \citep[see][for a description of an earlier version of 
this model]{fioc97}.  One can clearly see that, if one assumes a
universally-applicable
\citet{chabrier03} stellar IMF, that the star formation rate
and the growth of stellar mass paint a consistent picture, 
with around 40\% of all stellar mass being formed at $z < 1$.  

Yet, when one splits the galaxy population into red and 
blue galaxies, an interesting phenomenology emerges.
It is clear that blue galaxies contain much of the 
UV$+$IR-inferred star formation at every epoch.  If one assumes
that blue galaxies will stay blue at all times, one can integrate
the SFR to predict the growth of the stellar mass in blue cloud
galaxies: the result is shown in Fig.\ 2 as the dotted line.
It is clear that while the bulk of the stars are formed
in blue galaxies (as traced by the UV$+$IR), 
there is little growth in the total stellar mass in blue galaxies
with cosmic time.  Instead, the bulk of the growth in stellar 
mass is in red sequence (i.e., largely non-star-forming) galaxies.
{\it Blue, star-forming galaxies are shutting 
off their star formation on global scales and 
are fading onto the red sequence.}  

\section{Where is this mass transfer occurring?}    \label{diff}

Further insight into the evolution of stellar mass
in red and blue galaxies can be gained by 
examining the evolution of their stellar mass functions.
While such mass functions are well-defined only 
above $\sim 3 \times 10^{10} M_{\sun}$ (the mass limit at
$z \ga 0.6$), this exercise
gives some insight into the characteristics
of the galaxies which are transferred onto the 
red sequence from $z=1$ to the present-day.

In this section, we perform a simple analysis. 
One starts with the $z \sim 0.9$ stellar mass function
(of all, red and blue galaxies --- whatever population 
is of interest).  Then, one determines the average  
SFR in broad bins of stellar mass for the 
$0.8 < z \le 1.0$ galaxies (again, for all, red or blue
galaxies), and the average SFR per unit stellar
mass is determined as a function of stellar
mass (the specific SFR $dm/dt$; illustrated in the left-most set of panels
in Fig.\ \ref{fig:spsfr}).  Then, the average amount of 
stellar mass growth between $z = 0.9$ and $z=0.7$
is estimated, as a function of stellar mass, 
assuming a constant specific SFR in this interval and a locked-up 
mass fraction of 0.55 (i.e., 45\% of initially-formed
stellar mass is returned to the ISM; such a fraction 
is appropriate for a Chabrier 2003 or Kroupa 2001 IMF 
for populations with age $\sim 4$\,Gyr).  Functionally,
we adopt the simple approach of simply re-labelling the 
stellar mass bin labels of the stellar mass function with 
a new $z=0.7$ stellar mass $M(1+0.55 \Delta t\,dm/dt)$.
This predicted $z=0.7$ mass function (dashed line in the 2nd column
of panels in Figs.\ \ref{fig:hyp1} and \ref{fig:hyp2})
is then compared with the observed $z=0.7$ mass function
(solid lines).  The procedure is repeated between 
$z=0.7$ and $z=0.5$, and so on.  In this way, one 
then can estimate (in a very crude fashion) the 
evolution of the shape of the stellar mass function 
as a function of time, and ask if this prediction resembles
the observed evolution.

The result for all galaxies is shown in the uppermost panel of Figs.\ 
\ref{fig:hyp1} and \ref{fig:hyp2}.  Given the substantial systematic
uncertainties, and the crudeness of the analysis, the match between the 
predicted evolution of the stellar mass function and the observed
evolution is impressive.  One may notice that there is a slight
discontinuity in the behavior of the `all' and `red' stellar mass
functions from the $z \sim 0.3$ bin to the $z \sim 0$ bin.  This can 
be seen also in Fig.\ \ref{fig:integ}.  Part of this discontinuity 
reflects the methodology used for COMBO-17 mass estimation 
(see \S \ref{sec:mass}); part is likely attributable to remaining cosmic
variance at the low redshift end of COMBO-17.  

The lower panels of Fig.\ \ref{fig:hyp1} show the predicted evolution
of the stellar mass function of red and blue galaxies as a function 
of epoch, assuming that {\it all mass formed in blue galaxies
will stay in blue galaxies for all time, and all mass 
formed in red galaxies will stay in red galaxies} (i.e., no
transfer of mass between blue and red galaxies; the assumptions
used in constructing Fig.\ \ref{fig:integ}).  The
low specific SFRs of red-sequence galaxies lead to very modest
predicted mass growth, whereas a large amount is observed.
In contrast, the higher specific SFRs of blue-cloud galaxies 
lead to the formation of a large number of $\sim 10^{11} {\rm M}_{\sun}$
blue galaxies, in clear contradiction with the present-day blue cloud 
stellar mass function.  

Motivated by the overall similarity of the observed $z=0$ and $z\sim 0.9$ 
blue cloud stellar mass functions, we explore a second simple
hypothesis in Fig. \ref{fig:hyp2}: {\it all growth of the stellar
mass function is added to the red sequence} (i.e., the blue cloud
mass function is not allowed to change, and the red sequence
mass function is given all new galaxies produced in the overall mass
function)\footnote{It is also possible to carry out this exercise
by adding the galaxies created in the blue cloud and red
sequence separately. This method gives very similar results 
at the knee of the mass function, but overproduces slightly
the number of faint red sequence galaxies.  Such a discrepancy could
be addressed by arguing that massive blue galaxies truncate their
star formation more frequently than low mass galaxies.  
Yet, we do not overemphasize
challenges at the faint end, as behavior of the 
$z=0.9$ mass function at the faint end is observationally unconstrained.}.  
This hypothesis reproduces
the evolution of the blue and red stellar mass functions with 
striking accuracy (considering the significant systematic uncertainties
inherent to this kind of analysis).  

\section{Discussion}  \label{disc}

In this paper, we have performed a joint analysis
of galaxy stellar masses and star formation rates.  We found
an overall consistency between the predicted growth of 
stellar mass through star formation and the observed evolution 
of the stellar mass function (both in differential and 
integral forms).  We showed that the observed star formation 
in blue cloud galaxies is sufficient to dramatically overproduce
the present-day mass budget in blue galaxies.  Instead, a 
significant fraction of blue galaxies must have had their 
star formation suppressed by some physical process or processes, 
joining the red sequence.  While transformation of galaxies
may not be required at masses $\ga 10^{11} M_{\sun}$, large
numbers of galaxies with masses $\la 10^{11} M_{\sun}$ must
be transferred from the blue cloud to the red sequence at $z < 1$.

\subsection{The empirical basis for this paper}

While the analysis presented in this paper is clearly
model-dependent, the main results are qualitatively
robust.  The empirical basis for this statement is clear.
The rest-frame optical luminosity density in blue
cloud galaxies decreases by $\sim 1$\,mag since 
$z=1$; the change is to first order a fading in $L^*$
without accompanying change in $\phi^*$ 
\citep{lilly95,wolf03,willmer06}.  In this
interval, the rest-frame optical luminosity density
of red sequence galaxies stays roughly constant; 
the sense of this change is that $L^*$ fades by $\sim 1$\,mag
since $z=1$, accompanied by an increasing $\phi^*$ to keep the luminosity
density constant
\citep{chen03,bell04,faber06,brown06}.  The colors of 
both red and blue galaxies become redder with 
increasing time \citep{bell04,blanton06}; it is impossible
to avoid some increase in mass-to-light ratio 
with such aging \citep{belldejong,kauffmann03,bell03}.  
Thus, the stellar mass function of blue cloud 
galaxies must be approximately unchanging, 
while the stellar mass function of red sequence 
galaxies must undergo significant growth, at least
around $L^*$ \citep[see, e.g.,][]{bundy05,blanton06,borch}.  
On the other hand, UV, line emission and thermal-IR 
surveys all demonstrate convincingly that the bulk of 
star formation is in blue cloud, primarily disk-dominated
galaxies \citep{flores99,zheng04,hammer05,bell05,melbourne05}.  The 
IR emission, in particular, argues for significant extra
dust-obscured star formation above and beyond what is 
seen in the local Universe \citep{elbaz99,cha01,takeuchi05,lefloch05},
suggesting typical star formation rates for massive disks
of $\ga 10 M_{\sun}$\,yr$^{-1}$ (i.e., doubling times of order
a Hubble time; Hammer et al. 2005).  Thus, the massive blue
galaxies contain the bulk of the star formation; since one is not
allowed to change significantly the number of massive disk galaxies
(because of the luminosity function and color constraints), these 
rapidly-growing blue sequence galaxies must be balanced by a loss
to the (also rapidly building-up) red sequence.  In this way, despite
the many model dependencies --- the choice of 
bolometric correction for 24{\micron}-derived
star formation rates, calibrations of star formation 
rates and stellar mass-to-light ratios, choices
about faint-end slopes, choice of IMF, etc.\ --- it is 
clear that the overall picture presented in this paper is
rather robust.

\subsection{$3 \times 10^{10} M_{\sun}$: More than just the crossing point of mass functions?}

A particularly interesting result of this analysis 
is that the mass scale at which 
the bulk of the red sequence growth takes place is approximately the 
same as the mass scale at which the bulk of star formation occurs
on the blue cloud, at between $10^{10} M_{\sun}$ and $10^{11} M_{\sun}$;
i.e., at $\sim 3 \times 10^{10} M_{\sun}$.
It was not obvious at all that this needed to be so.
We ourselves argued in \citet{bell04} that there were an insufficient
number of massive disk galaxies with which to feed the red mass function 
growth: faced with this mismatch, we appealed to widespread and frequent
merging of low mass $\sim 10^{10} M_{\sun}$ galaxies to produce more massive
red sequence counterparts.
Such a view is manifestly mistaken, however.  
Given the results from ISO and Spitzer, it is now clear that
massive disk galaxies host a large amount of star formation, sufficient
to double their mass in a Hubble time or less 
\citep{hammer05,melbourne05,bell05}.  Such intense star formation, 
obscured by dust, is sufficient to dramatically overproduce the 
number of local massive disks.  This shifts our perception of the 
blue cloud significantly. A large fraction of massive blue galaxies
are {\it required} to quench their star formation and join the 
red sequence; this loss of massive blue galaxies is compensated
for by growth of lower-mass blue cloud galaxies through intense 
star formation.

In the past, much significance has been attached
to the transition in galaxy properties
seen at $\sim 3 \times 10^{10} M_{\sun}$, presented
first (in a compelling form) by \citet{kauffmann03b}. 
In its original incarnation, this transition is 
defined as the mass at which blue galaxies and red galaxies
are equally common, i.e., where the blue and red galaxy stellar
mass functions cross.  Subsequent work has found other
galaxy properties to change at this transition mass, e.g., 
AGN activity is much more common above this mass than 
below it \citep{kauffmannagn}.  Recent works have probed
the evolution of the cross-over of the red and blue mass functions
out to $z=1$, finding an order of magnitude decrease in the 
cross-over point from $z=1$ to the present day 
\citep{bundy05,borch}.  
We take the position
that this cross-over point is to a certain extent a historical
happenstance; indeed, it is observed to be a strong function of 
environment in the local Universe \citep{baldry06}.  Instead, 
we attach much more significance to the cut-off of the blue 
cloud stellar mass function, starting at $\sim 3 \times 10^{10} M_{\sun}$.
This cut-off, which on the basis of this analysis we interpret 
as an important scale for global suppression of star formation 
and transformation to red sequence galaxies, is {\it not} a strong 
function of redshift or environment.

\subsection{Expectations of this picture}

There are three predictions/expectations of this approach
which can be compared with independent analyses.
Firstly, it is increasingly apparent from previous analyses as
well as our own that massive red sequence
galaxies with $M_* > 10^{11} M_{\sun}$ 
should have older stellar population than those at $3 \times 10^{10}$
\citep[see, e.g.,][]{juneau05,cimatti06,panter06}.
On the basis of the observed $b \sim 1$ of blue cloud galaxies
(galaxies whose SFR has not yet been suppressed) and the observed
growth of the red sequence, one might expect that
the average age of a red sequence galaxy 
with $3 \times 10^{10} M_{\sun}$ should be $\la 8$\,Gyr.  
At masses $> 2 \times 10^{11} M_{\sun}$, 
little growth is observed; one would expect most star formation 
to happen at $z >1$ \citep[as is indeed observed;][]{papovich06,daddi06}, 
giving typical `ages'
of $\sim 8-12$\,Gyr.  This compares favorably
with observations of the stellar population ages of 
early-type galaxies from \citet{thomas05}, discussed also by 
\citet{thomas06}: galaxies with $\sim 3 \times 10^{10} M_{\sun}$
have ages between 5 and 9\,Gyr \citep[adopting the age scatter
of $\sim 0.25t$, where $t$ is the age, from Thomas et al.\ 2005]{thomas06}, 
while galaxies with masses   $> 2 \times 10^{11} M_{\sun}$ 
have ages $9-13$\,Gyr \citep{thomas06}.  

Secondly, on a related theme, one expects also a relatively prominent 
population of galaxies undergoing truncation of their star formation 
at masses between $10^{10} M_{\sun}$ and $10^{11} M_{\sun}$.  
Such a population will be characterized by a rapidly-declining or 
truncating star formation rate; depending on the rapidity of the 
truncation, and on the distribution of dust in the galaxy, one is likely
to see some spectral evidence of the rapid decrease star formation 
rate (e.g., enhanced Balmer lines and weak/nonexistent emission lines).
In this context, the observations of \citet{leborgne06} are particularly
interesting.  They study galaxies with 
$M > 1.7 \times 10^{10} M_{\sun}$ at $0.6< z < 1.2$ and the present day 
using the Gemini Deep Deep Survey and the Sloan Digital Sky Survey, finding
a decrease in the ``H$\delta$ strong'' fraction from $\sim$50\% at 
$z \sim 1.2$ to a few percent at present.  Extension of 
this methodology to larger samples (allowing splitting into different
mass bins) will allow a more detailed investigation of the 
redshift evolution, environmetal dependence, and mass dependence
of the truncation of star formation in galaxies.

Finally, depending on transformation mechanism, one may expect
quite a bit of environmental dependence in galaxy ages; exploration
of this issue is well beyond
the scope of this work.  Qualitatively, however, 
given that in this picture a large fraction of current red galaxies
are generated through suppression of star formation in galaxies which 
were blue at $z \sim 1$, and that red galaxies currently live in 
reasonably dense environments, one expects that a large number
of massive blue $z \sim 1$ galaxies will be relatively strongly clustered ---
clustered much more strongly than present-day massive blue galaxies.
Such an expectation is spectacularly borne out by the data:
analyses of both the DEEP2 and VVDS galaxy evolution 
surveys demonstrate an abundance of unusually strongly-clustered
massive blue galaxies \citep{cooper06,cucciati06}.  For workers wishing to 
understand the driving forces of early-type galaxy evolution, 
these galaxies merit special attention: analyses of their
environment, morphological and structural properties, stellar
populations, and supermassive black hole content will shed light
on these `proto-early-type' galaxies.

\section{Conclusions}

In this paper, we have presented a joint analysis of galaxy 
stellar masses and star formation rates from the COMBO-17 
photometric redshift survey combined with Spitzer 24{\micron}
data.  Stellar masses were estimated using 17-passband optical
spectral energy distribution fitting, and star formation rates
were determined from the combined ultraviolet and IR fluxes of galaxies.
We estimated SFR functions for all galaxies in 4 redshift bins, and presented
for the first time SFR functions split into contributions from 
blue cloud and red sequence galaxies.

When standard stellar mass and SFR calibrations were used, we found 
an overall consistency between the observed growth of stellar mass
since $z=1$ and the integral of the cosmic SFR.  Yet, the bulk of 
SFR occurs in blue cloud galaxies; if all stars either already in or 
formed in blue galaxies since $z=1$ were to end up in present-day blue cloud 
galaxies, the total stellar mass budget in blue cloud galaxies today
would be dramatically overproduced by a factor of two.  
Instead, the stellar mass density
in red sequence galaxies grows steadily from $z=1$ to the present 
by approximately the amount required to balance the `overproduction'
of stars in blue cloud galaxies.  Thus, a large fraction of 
blue galaxies must have their star formation suppressed
by some physical process or processes, joining the red sequence.

We explored the specific SFRs (i.e., SFR per unit
stellar mass) as a function of galaxy mass.  These specific SFRs 
were calculated using stacking of 24{\micron} data on the positions
of the galaxies of interest, in combination 
with their rest-frame UV luminosities from COMBO-17.  Using these
specific SFRs, we `predicted' the evolution of the stellar mass function
from one redshift bin to the next.  We found that the evolution of the
overall stellar
mass function is reasonably well reproduced using this approach. 
Yet, if one assumes that all stars already in or formed in
blue galaxies since $z<1$ stay in blue galaxies, one dramatically
overproduces the number of present-day {\it massive} blue galaxies 
and underproduces the growth of red sequence galaxies 
at $\la 10^{11}M_{\sun}$.  This predicted excess of massive blue 
galaxies is approximately sufficient to feed the growth of the red
sequence at $\la 10^{11}M_{\sun}$.   Thus, not only
must a significant fraction of blue star forming galaxies suppress
their star formation between $z=1$ and the present day, but also
the observations require that the suppression mechanism or mechanisms
be particularly effective at $\sim 3 \times 10^{10} M_{\sun} $.

\acknowledgements
We are thankful for a constructive and thoughtful referee's report, which
led to significant improvement of the paper.
E.\ F.\ B.\ was supported by the European Community's Human
Potential Program under contract HPRN-CT-2002-00316 (SISCO)
and by the Emmy Noether Programme of the Deutsche
Forschungsgemeinschaft.
C.\ P.\ was supported by NASA through the {\it Spitzer 
Space Telescope} Fellowship Program, through a contract issued 
by the Jet Propulsion Laboratory, California Institute of Technology
under a contract with NASA.
C.\ W.\ was supported by a PPARC Advanced Fellowship.


\begin{thebibliography}{}

\bibitem[Alonso-Herrero et al.(2006)]{alonso06}
	Alonso-Herrero, A., et al. 2006, \apj, 640, 167

\bibitem[Appleton et al.(2004)]{appleton04}
	Appleton, P.\ N., et al. 2004, \apjs, 154, 147

\bibitem[Baldry et al.(2006)]{baldry06}
  Baldry, I.\ K., Balogh, M., Bower, R., Glazebrook, K., 
  Nichol, R., Bamford, S., \& Budavari, T. 2006, \mnras, 373, 469

\bibitem[Bell \& de Jong(2001)]{belldejong}
	Bell, E.\ F., \& de Jong, R.\ S. 2001, \apj, 550, 212

\bibitem[Bell(2003)]{bellsfr}
	Bell, E.\ F. 2003, \apj, 586, 794

\bibitem[Bell et al.(2003)]{bell03}
	Bell, E.\ F., McIntosh, D.\ H., Katz, N., \& Weinberg, M.\ D. 2003, 
	\apjs, 149, 289

\bibitem[Bell et al.(2004)]{bell04}
	Bell, E.\ F., et al. 2004, \apj, 608, 752

\bibitem[Bell et al.(2005)]{bell05}
	Bell, E.\ F., et al. 2005, \apj, 625, 23

\bibitem[Blanton et al.(2003)]{blanton03}
        Blanton, M.\ R., et al. 2003, \apj, 594, 186

\bibitem[Blanton(2006)]{blanton06}
        Blanton, M.\ R. 2006, \apj, 648, 268

\bibitem[Borch et al.(2006)]{borch}
	Borch, A., Meisenheimer, K., Bell, E.\ F., 
	Rix, H.-W., Wolf, C., Dye, S., Kleinheinrich, M., \& Kovacs, Z.
	2006, \aap, 453, 869

\bibitem[Brand et al.(2006)]{brand06}
	Brand, K., et al. 2006, \apj, 644, 143

\bibitem[Brinchmann \& Ellis(2000)]{brinchmann00}
        Brinchmann, J., \& Ellis, R.\ S. 2000, \apj, 536, 77L

\bibitem[Brown et al.(2006)]{brown06}
  Brown, M.\ J.\ I., Dey, A., Jannuzi, B.\ T., Brand, K., Benson, A.\ J.,
	Brodwin, M., Croton, D.\ J., \& Eisenhardt, P.\ R. 2007, \apj, 654, 858

\bibitem[Bundy et al.(2006)]{bundy05}
	Bundy, K., et al. 2006, \apj, 651, 120

\bibitem[Chabrier(2003)]{chabrier03}
        Chabrier, G. 2003, ApJ, 586, L133

\bibitem[Chary \& Elbaz(2001)]{cha01}
Chary, R., \& Elbaz, D. 2001, \apj, 556, 562

\bibitem[Chen et al.(2003)]{chen03}
        Chen, H.-W., et al. 2003, \apj, 586, 745

\bibitem[Cimatti, Daddi, \& Renzini(2006)]{cimatti06}
	Cimatti, A., Daddi, E., \& Renzini, A. 2006, \aap, 453, L29

\bibitem[Cole et al.(2001)]{cole01}
        Cole, S., et al. 2001, \mnras, 326, 255

\bibitem[Cooper et al.(2006)]{cooper06}
	Cooper, M., et al. 2006, \mnras, 370, 198

\bibitem[Cucciati et al.(2006)]{cucciati06}
	Cucciati, O., et al. 2006, \aap, 458, 39

\bibitem[Daddi et al.(2005)]{daddi06}
  Daddi, E., et al. 2005, \apj, 631, L13

\bibitem[Dale et al.(2001)]{dale01}
Dale, D.\ A., Helou, G., Contursi, A., Silbermann, N.\ A., \& Kolhatkar,
S.\ 2001, \apj, 549, 215

\bibitem[Devriendt et al.(1999)]{dev99}
Devriendt, J.\ E.\ G., Guiderdoni, B., Sadat, R.\ 1999, \aap, 350, 381

\bibitem[Dickinson et al.(2003)]{dickinson03}
Dickinson, M., Papovich, C., Ferguson, H.\ C., \& Budav\'ari, T. 2003, 
\apj, 587, 25

\bibitem[Donley et al.(2005)]{donley05}
  Donley, J.\ L., Rieke, G.\ H., Rigby, J.\ R., P\'erez-Gonz\'alez, 
  P.\ G. 2005, \apj, 634, 169

\bibitem[Drory et al.(2004)]{drory04}
	Drory, N., Bender, R., Feulner, G., Hopp, U., Maraston, C., 
	Snigula, J., \& Hill, G.\ J. 2004, \apj, 608, 742

\bibitem[Drory et al.(2005)]{drory05}
Drory, N., Salvato, M., Gabasch, A., Bender, R., Hopp, U., Feulner, G.,
\& Pannella, M. 2005, \apj 619, L131

\bibitem[Elbaz et al.(1999)]{elbaz99}
        Elbaz, D., et al. 1999, \aap, 351, L37

\bibitem[Elbaz et al.(2005)]{elbaz05}
	Elbaz, D., Le Floc'h, E., Dole, H., \& Marcillac, 
	D. 2005, \aap, 434, L1

\bibitem[Elmegreen(2006)]{elmegreen06}
  Elmegreen, B.\ G. 2006, \apj, 648, 572

\bibitem[Faber et al.(2006)]{faber06}
	Faber, S.\ M., et al. 2006, submitted to {\apj} (astro-ph/0506044)

\bibitem[Fioc \& Rocca-Volmerange(1997)]{fioc97}
	Fioc, M., \& Rocca-Volmerange, B. 1997, \aap, 326, 950

\bibitem[Flores et al.(1999)]{flores99}
        Flores, H., et al. 1999, \apj, 517, 148

\bibitem[Fontana et al.(2004)]{fontana}
        Fontana, A., et al. 2004, \aap, 424, 23

\bibitem[Fontana et al.(2006)]{fontana06}
        Fontana, A., et al. 2006, \aap, 459, 745

\bibitem[Franceschini et al.(2005)]{fran05}
        Franceschini, A., et al. 2005, \aj, 129, 2074

\bibitem[Gordon et al.(2000)]{fluxrat}
        Gordon, K.\ D., Clayton, G.\ C., Witt, A.\ N., Misselt, K.\ A. 2000,
                \apj, 533, 236

\bibitem[Gordon et al.(2005)]{dat}
        Gordon, K.\ D., et al. 2005, \pasp, 117, 503

\bibitem[Haarsma et al.(2000)]{haarsma00}
        Haarsma, D.\ B., Partridge, R.\ B., Windhorst, R.\ A., \&
        Richards, E.\ A. 2000, \apj, 544, 641

\bibitem[Hammer et al.(2005)]{hammer05}
  Hammer, F., Flores, H., Elbaz, D., Zheng, X.\ Z., Liang, Y.\ C., \& 
  Cesarsky, C. 2005, \aap, 430, 115

\bibitem[Helou et al.(1988)]{helou88}
        Helou, G., Khan, I.\ R., Malek, L., Boemher, L. 1988, \apjs, 68, 151

\bibitem[Hopkins(2004)]{hopkins04}
        Hopkins, A.\ M., 2004, \apj, 615, 219

\bibitem[Juneau et al.(2005)]{juneau05}
	Juneau, S., et al. 2005, \apj, 619, L135

\bibitem[Kauffmann et al.(2003a)]{kauffmann03}
        Kauffmann, G., et al. 2003a, \mnras, 341, 33

\bibitem[Kauffmann et al.(2003b)]{kauffmann03b}
        Kauffmann, G., et al. 2003b, \mnras, 341, 54

\bibitem[Kauffmann et al.(2003c)]{kauffmannagn}
        Kauffmann, G., et al. 2003c, \mnras, 341, 1055

\bibitem[Kennicutt(1998)]{k98}
        Kennicutt Jr., R.\ C. 1998, \araa, 36, 189

\bibitem[Kroupa(2001)]{kroupa01}
        Kroupa, P. 2001, \mnras, 322, 231

\bibitem[Kroupa et al.(1993)]{kroupa93}
Kroupa, P., Tout, C.\ A., \& Gilmore, G. 1993, \mnras, 262, 545

\bibitem[Le Borgne et al.(2006)]{leborgne06}
	Le Borgne, D., et al. 2006, \apj, 642, 48

\bibitem[Le F\`evre et al.(2000)]{lefevre00}
	Le F\`evre, et al. 2000, \mnras, 311, 565

\bibitem[Le Floc'h et al.(2004)]{lefloch04}
        Le Floc'h, E., et al. 2004, \apjs, 154, 170

\bibitem[Le Floc'h et al.(2005)]{lefloch05}
        Le Floc'h, E., et al. 2005, \apj, 632, 169

\bibitem[Lilly et al.(1995)]{lilly95}
        Lilly, S.\ J., Tresse, L., Le F\`evre, O., Hammer, F., \&
        Crampton, D. 1995,
        \apj, 455, 75

\bibitem[Lilly et al.(1996)]{lilly96}
        Lilly, S.\ J., Le F\`evre, O., Hammer, F., Crampton, D. 1996,
        \apj, 460, L1

\bibitem[Lin et al.(2005)]{lin05}
	Lin, L., et al. 2005, ApJ, 671, L9

\bibitem[Madau et al.(1996)]{madau96}
        Madau, P., Ferguson, H.\ C., Dickinson, M.\ E., Giavalisco, M.,
        Steidel, C.\ C., \& Fruchter, A. 1996, \mnras, 283, 1388

\bibitem[Mart\'{\i}nez-Sansigre et al.(2006)]{martinez06}
  Mart\'{\i}nez-Sansigre, A., Rawlings, S., Lacy, M., Fadda, D., 
  Jarvis, M.\ J., Marleau, F.\ R., Simpson, C., Willott, C.\ J. 
  2006, \mnras, 370, 1479

\bibitem[Melbourne et al.(2005)]{melbourne05}
  Melbourne, J., Koo, D.\ C., \& Le Floc'h, E. 2005, \apj, 632, L65

\bibitem[Panter et al.(2006)]{panter06}
	Panter, B., Jiminez, R., Heavens, A.\ F., \& Charlot, S. 2006, 
	submitted to {\mnras} (astro-ph/0608531)

\bibitem[Papovich \& Bell(2002)]{papovich02}
	Papovich, C., \& Bell, E.\ F. 2002, \apj, 579, L1

\bibitem[Papovich et al.(2004)]{papovich04}
	Papovich, C., et al. 2004, \apjs, 154, 70

\bibitem[Papovich et al.(2006)]{papovich06}
	Papovich, C., et al. 2006, \apj, 640, 92

\bibitem[Patton et al.(2002)]{patton02}
	Patton, D.\ R., et al. 2002, \apj, 565, 208

\bibitem[Roussel et al.(2001)]{rou01}
Roussel, H., Sauvage, M., Vigroux, L., \& Bosma, A.\ 2001, \aap, 372, 427

\bibitem[Rudnick et al.(2003)]{rudnick03}
Rudnick G., et al. 2003, \apj, 599, 847

\bibitem[Rudnick et al.(2006)]{rudnick06}
Rudnick G., et al. 2006, \apj, 650, 624

\bibitem[Salpeter(1955)]{salp}
        Salpeter, E.\ E. 1955, \apj, 121, 161

\bibitem[Silva, Maiolino, \& Granato(2004)]{silva04}
        Silva, L., Maiolino, R., \& Granato, G.\ L. 2004, \mnras,
        355, 973

\bibitem[Spinoglio et al.(1995)]{spi95}
Spinoglio, L., Malkan, M.\ A., Rush, B., Carrasco, L., \&
Recillas-Cruz, E. 1995, \apj, 453, 616

\bibitem[Steidel et al.(1999)]{steidel99}
        Steidel, C.\ C., Adelberger, K.\ L., Giavalisco, M., Dickinson, M., \&
        Pettini, M. 1999, \apj, 519, 1

\bibitem[Strateva et al.(2001)]{strateva01}
        Strateva, I., et al. 2001, \aj, 122, 1861

\bibitem[Takeuchi, Buat \& Burgarella(2005)]{takeuchi05}
  Takeuchi, T.\ T., Buat, V., \& Burgarella, D. 2005, \aap, 440, L17

\bibitem[Thomas \& Davies(2006)]{thomas06}
  Thomas, D., \& Davies, R.\ L. 2006, \mnras, 366, 510

\bibitem[Thomas et al.(2005)]{thomas05}
  Thomas, D., Maraston, C., Bender, R., \& Mendes de Oliveira, C. 2005, 
  \apj, 621, 673

\bibitem[Willmer et al.(2006)]{willmer06}
	Willmer, C., et al. 2006, \apj, 647, 853

\bibitem[Wolf et al.(2003)]{wolf03}
	Wolf, C., Meisenheimer, K., Rix, H.-W., Borch, A., Dye, S., \&
	Kleinheinrich, M. 2003, \aap, 401, 73

\bibitem[Wolf et al.(2004)]{wolf04}
	Wolf, C., et al. 2004, \aap, 421, 913

\bibitem[Wolf et al.(2005)]{wolf05}
	Wolf, C., et al. 2005, \apj, 630, 771

\bibitem[Yan et al.(2005)]{yan05}
	Yan, L., et al. 2005, 2005, \apj, 628, 604

\bibitem[Zheng et al.(2004)]{zheng04}
        Zheng, X.\ Z., Hammer, F., Flores, H., Ass\'emat, F., \&
        Pelat, D. 2004, \aap, 421, 847

\bibitem[Zheng et al.(2006)]{zheng06}
        Zheng, X.\ Z., et al. 2006, \apj, 640, 784

\bibitem[Zheng et al.(2007a)]{zheng07_70160}
        Zheng, X.\ Z., Dole, H., Bell, E.\ F., Le Floc'h, E., 
	Rieke, G.\ H., Rix, H.-W., \& Schiminovich, D. 
	2007a, submitted to {\apj}

\bibitem[Zheng et al.(2007b)]{zheng07_down}
        Zheng, X.\ Z., et al.
	2007b, {\apj} Letters, in press (astro-ph/0702208)


\end{thebibliography}
\end{document}